\documentclass{article}

\usepackage{amssymb,amsfonts,amsmath,stmaryrd,graphicx,calc}
\usepackage{cite,enumerate,float,indentfirst}
\usepackage{color}

\def\be{\begin{eqnarray}}
\def\ee{\end{eqnarray}}
\def\nn{\nonumber}

\def\Gns{{\rm Gns}}

\newcommand{\beq}{\begin{equation}}
\newcommand{\eeq}{\end{equation}}
\newcommand{\beqa}{\begin{eqnarray}}
\newcommand{\eeqa}{\end{eqnarray}}

\newcommand{\lam}{\lambda}
\newcommand{\m}{\mu}

\definecolor{red}{rgb}{1,0,0}
\definecolor{orange}{rgb}{1,0.5,0}
\definecolor{violet}{rgb}{0.7,0,1}

\hoffset= - 1.0in         

\begin{document}

\title{
Shiraishi functor and non-Kerov deformation of Macdonald polynomials
}

\author{
{\bf  Hidetoshi Awata$^a$}\footnote{awata@math.nagoya-u.ac.jp}, \
{\bf Hiroaki Kanno$^{a,b}$}\footnote{kanno@math.nagoya-u.ac.jp}, \
{\bf Andrei Mironov$^{c,d,e}$}\footnote{mironov@lpi.ru; mironov@itep.ru},
 and \  {\bf Alexei Morozov$^{f,d,e}$}\thanks{morozov@itep.ru}
\date{ }
}

\maketitle

\vspace{-6.0cm}

\begin{center}
\hfill FIAN/TD-02/20\\
\hfill IITP/TH-02/20\\
\hfill ITEP/TH-02/20\\
\hfill MIPT/TH-02/20
\end{center}

\vspace{4.7cm}

\begin{center}
$^a$ {\small {\it Graduate School of Mathematics, Nagoya University, Nagoya, 464-8602, Japan}}\\
$^b$ {\small {\it KMI, Nagoya University, Nagoya, 464-8602, Japan}}\\
$^c$ {\small {\it Lebedev Physics Institute, Moscow 119991, Russia}}\\
$^d$ {\small {\it ITEP, Moscow 117218, Russia}}\\
$^e$ {\small {\it Institute for Information Transmission Problems, Moscow 127994, Russia}}\\
$^f$ {\small {\it MIPT, Dolgoprudny, 141701, Russia}}
\end{center}

\vspace{.0cm}

\begin{abstract}
We suggest a further generalization of the hypergeometric-like series due to M. Noumi and J. Shiraishi
by substituting the Pochhammer symbol with a nearly arbitrary function. Moreover, this generalization is valid for
the entire Shiraishi series, not only for its Noumi-Shiraishi part.
The theta function needed in the recently suggested description of the double-elliptic systems \cite{AKMM2},
6d N=2* SYM instanton calculus and the doubly-compactified network models,
is a very particular member of this huge family.
The series depends on two kinds of variables, $\vec x$ and $\vec y$,
and on a set of parameters, which becomes infinitely large now.
Still, one of the parameters, $p$ is distinguished by its role in the series grading.
When $\vec y$ are restricted to a discrete subset labeled by Young diagrams,
the series multiplied by a monomial factor reduces to a polynomial at any given order in $p$.
All this makes the  map from functions to the hypergeometric-like series very promising,
and we call it {\it Shiraishi functor}
despite it remains to be seen, what are exactly the morphisms that it preserves.
Generalized Noumi-Shiraishi (GNS) symmetric polynomials inspired by the Shiraishi functor
in the leading order in $p$ can be obtained by
a triangular transform from the Schur polynomials and possess an interesting grading.
They provide a family of deformations of Macdonald polynomials,
as rich as the family of Kerov functions,
still very different from them, and, in fact, much closer to the Macdonald polynomials.
In particular, unlike the Kerov case, these polynomials do not depend on the ordering of
Young diagrams in the triangular expansion.
\end{abstract}

\section{Introduction}

In the present paper, we discuss a new insight into the theory of
generalized hypergeometric series $\mathfrak{P}[\vec x|\vec y]$ generalizing those
introduced by J. Shiraishi \cite{S}.
Here $\vec y$ parameterize the moduli in the mother-function formalism (see \cite{AKMM1} for a recent review),
and the partitions, or Young diagrams $R$ (on which the standard symmetric polynomials like Macdonald depend)
appear on the particular locus
\be
y_i = q^{R_i}(st)^{N-i}
\label{ydia0}
\ee
Our main goal is to demonstrate that the construction can and deserves
to be further extended, and provides a series $\mathfrak{P}_\xi[\vec x|\vec y]$
for {\it any} input function $\xi(z)$,
restricted by two simple conditions
\be
\xi(1)=0 \ \ \ \ {\rm and}  \ \ \ \ \xi(z^{-1}) \sim z^{-1} \xi(z)
\ee
Then, the Shiraishi series {\it per se} is associated with $\xi(z) =1-z$,
while the elliptic deformation, supposedly relevant \cite{AKMM2} to the double-elliptic systems
\cite{MMdell}, is associated with elliptic $\xi(z) \sim \vartheta(z)$ (see also \cite{FOS} for the elliptic case).
However, at each power of $p$ we get  polynomials in $x$
on the locus (\ref{ydia0}) for arbitrary $\xi(z)$.
Moreover, at least in the leading order in $p$, these polynomials are as nice as the ordinary Macdonald polynomials:
their symmetric part is obtained by a triangular transformation from the symmetric polynomials ${\rm Mon}_R[x]$
and should satisfy a Calogero/Ruijsenaars-like equations
(i.e. it is expected to be possible to define an appropriate $\xi$-dependent version of
the Calogero-Ruijsenaars Hamiltonians and integrable systems).
We call this entire construction {\it Shiraishi functor}, which is
hopefully acting from the space of functions $\xi(z)$ into that of integrable systems.

In what follows, for each function $\xi(z)$ we introduce the series that depends on variables
$x_i$, $y_i$, $i=1,\ldots,N$, which is a formal power series in the ratios $x_i/x_j$
in sec.\ref{GenShirdef}.
After a certain rescaling Shiraishi series becomes also a series in positive powers of $p^N$.
Moreover, being multiplied by a proper monomial $\prod_i x_i^{R_i}$,
every item becomes a polynomial in $x_i$ after the substitution (\ref{ydia0})
is made for $\vec y$.
This completes the definition of {\it Shiraishi functor}.

In the rest of the paper we mostly concentrate on {\it generalized Noumi-Shiraishi} (GNS)
polynomials, which appear in the order $p^0$ after symmetrization (see \cite{KS} for $N=2$ GNS in the particular elliptic case).
The set of such symmetric polynomials is actually as big as another well known family
of deformations of Macdonald polynomials: that of Kerov functions
also labeled by an arbitrary functions $g(z) = \sum_{i=1}^\infty g_iz^i$.
Still the Shiraishi set is completely different from the Kerov one, and they intersect only over
the Macdonald polynomials.
Moreover, as we explain, the GNS polynomials have a good chance to preserve
close links to representation theory, preserved by the Macdonald deformation of the Schur polynomials,
but violated by the Kerov functions.

Our claim is that, depending on entire arbitrary function $\xi(z)$,
i.e. on infinitely many additional parameters,
they remain very similar to Macdonald polynomials, though very different from Kerov functions.
They are obtained by a triangular transform from monomial symmetric polynomials, or from Schur polynomials,
or from Macdonald polynomials, and thus can be easily continued from the Miwa locus
to the entire space of time variables $p_k$.
The coefficients actually depend on $\xi$ through a peculiar function
$\eta(z) = \frac{\xi(qz)\xi(t/qz)}{\xi(tz)\xi(z)}$,
moreover, in the basis of monomial symmetric polynomials there is a simple grading,
connecting the power of $\eta$ and the number of monomial symmetric polynomials 
in the lexicographical ordered list of partitions at a given level.
For the complementary system, conjugate to the GNS polynomials w.r.t. the Schur measure,
the grading also shows up in a proper basis.
With the help of this conjugate system, one can build up the Cauchy formula
and define the skew GNS polynomials
through the corresponding generalized Littlewood-Richardson coefficients. We also manage to construct a Hamiltonian
that has the GNS polynomials associated with the one-row partitions as its eigenfunctions.
In variance with the Kerov case, in the definition of the GNS polynomials, it is sufficient to choose the normal (partial) ordering,
and, hence, the intersection of the sets of GNS and Kerov polynomials
is exhausted by the Macdonald polynomials only.

\paragraph{Notation.}

In the paper, we denote the polynomials:
\be
\begin{array}{cl}
{\rm Schur}&\hbox{Schur polynomials \cite{Mac}}\\
{\rm Mac}&\hbox{Macdonald polynomials \cite{Mac}}\\
{\rm Mon}&\hbox{monomial symmetric polynomials \cite{Mac}}\\
{\rm Kerov}&\hbox{Kerov functions \cite{Kerov,MMkerov}}\\
{\rm Gns},\ {\rm Gns}^\perp&\hbox{generalized Noumi-Shiraishi polynomials (\ref{GNS}), (\ref{ex2})}\\
p_k:=\sum_ix_i^k&\hbox{power symmetric polynomials}
\end{array}
\ee
For the Young diagram $R=\{R_1\geq R_2 \geq \ldots\}$, we use the notations $p_R:=\prod_i p_{R_i}$ and
$z_R:=\prod_k k^{m_k}m_k!$ where $m_k$ is the number of lines of length $k$ in the diagram $R$. We use the functions
\be
\eta(z) = \frac{\xi(qz)\xi(t/qz)}{\xi(tz)\xi(z)}\nn\\
\zeta_k(z) =   \frac{\xi(q^kz)\xi(tz)}{\xi(q^{k-1}tz)\xi(qz)} = \prod_{i=1}^{k-1} \eta(q^iz)
\ee
Throughout the paper, we call {\bf Macdonald locus} the choice of function $\xi(z)=1-z$. 

\section{Shiraishi functor: promoting Macdonald $1-z$ to arbitrary function
\label{GenShirdef}}

Now we give a preliminary definition of {\it Shiraishi functor}
that allows one to convert an arbitrary function $\xi(z)$ into a map from a power series of $2N$ variables
parameterized by the function $\xi(z)$ and by four parameters
to an infinite set of graded functions parameterized by the Young diagrams $R$.

Suppose we are given a function $\xi(z)$ with the symmetry properties
\be
\xi(1)=0,\ \ \ \ \ \xi(z^{-1})=\alpha z^{-1}\xi(z),\ \ \ \ \alpha\in\mathbb{C}
\ee
Define
\beq\label{Theta}
\Xi(z;q)_n =
\begin{cases}
\displaystyle{\prod_{k=0}^{n-1}}\xi(q^k z), \qquad n \geq 0, \\
\displaystyle{\prod_{k=0}^{n-1}} \xi (q^{-k-1} z)^{-1}, \qquad n < 0.
\end{cases}
\eeq
where $|q|<1$ is a parameter. Now one can define a $\xi$-Shiraishi power series
\beq
\mathfrak{P}_N^{\xi} { (x_i ; p \vert y_i ; s \vert q,t)}
:= \sum_{\lam^{(i)}} \prod_{i,j=1}^N
\frac{\mathcal{N^\xi}_{\lam^{(i)}, \lam^{(j)}}^{(j-i)} (t y_j/y_i \vert q,s)}
{\mathcal{N^\xi}_{\lam^{(i)}, \lam^{(j)}}^{(j-i)} (y_j/y_i \vert q,s)}
\prod_{\beta=1}^N \prod_{\alpha \geq 1}
\left( \frac{p x_{\alpha + \beta}}{t x_{\alpha + \beta -1}} \right)^{\lam_\alpha^{(\beta)}}
\label{EG}
\eeq
where
\beqa \label{EGfactor}
\mathcal{N^\xi}_{\lam, \mu}^{(k)} (u \vert q, s)
&:=& \!\!\!\!\!\!\!\!
\prod_{b \geq a \geq 1 \atop b-a \equiv k~(\mathrm{mod}~n)}
\!\!\!\!\!\!\!\!
\Xi\Big(uq^{-\mu_a + \lam_{b+1}} s^{-a +b} ; q\Big)_{_{\lam_b - \lam_{ b+1}}}
\times\!\!\!\!\!
\prod_{b \geq a \geq 1 \atop b  - a \equiv -k-1~(\mathrm{mod}~n)}
\!\!\!\!\!\!\!\!
\Xi \Big(uq^{\lam_a - \m_b} s^{a - b -1} ; q\Big)_{_{\m_b - \m_{b+1}}}
\eeqa
where $\{\lambda^{(i)}\}$, $i=1,\ldots,N$ is a set of $N$ partitions, and we assume $x_{N+i}=x_i$. Then,
\beq
{\bf P}^\xi_R(x_i;p,s,q,t):=\prod_{i=1}^Nx_i^{R_i}\cdot
\mathfrak{P}_N^{\xi} { \left(p^{N-i}x_i ; p\, \Big|\, y_i=q^{R_i}(st)^{N-i} ; s \,\Big|
\, q,\frac{q}{t}\,\right)}
\eeq
is a graded function of variables $x_i$ of the weight $|R|=\sum_iR_i$,
which is a series in $p^{Nk}$:
\be
{\bf P}^\xi_R(x_i;p,s,q,t)=\sum_{k\ge 0}p^{Nk}\cdot{{\bf P^\xi}^{(k)}_R(x_i;s,q,t)\over \prod_{i=1}^N x_i^k}=
\sum_{k\ge 0}{\bf P^\xi}^{(k)}_R(x_i;s,q,t)\cdot\prod_{i=1}^N\left({p\over x_i}\right)^k
\ee
Here ${\bf P^\xi}^{(k)}_R(x_i;s,q,t)$ is a polynomial of variables $x_i$ with grade $|R|+Nk$.

Note that, in (\ref{EGfactor}), instead of using the standard Nekrasov function, we impose the mod $n$ selection rule following \cite{S}.
It is essential for assuring ${\bf P^\xi}^{(k)}_R(x_i;s,q,t)$ to be a polynomial, moreover, to be a symmetric polynomial at $N=2$ and almost a symmetric polynomial at larger $N$ (see the next section).

If one chooses $\xi(z)=1-z$, then
$\mathfrak{P}_N^{\xi} { (x_i ; p \vert y_i ; s \vert q,t)}$ is the standard Shiraishi function.
Choosing $\xi(z)=\sqrt{z}\vartheta_p(z)/(p;p)_\infty$,
one gets an elliptic lift of the Shiraishi function. In the both these cases, $\alpha=-1$.

\section{Symmetric polynomials}

The main problem with this definition is that defined in {\it such} a way ${\bf P^\xi}^{(k)}_R(x_i;s,q,t)$
are not obligatory symmetric in $x_i$ (beyond the Macdonald case).
This may imply that the definition should be somehow modified.
In what follows, we use a very concrete prescription to make symmetric polynomials of ${\bf P^\xi}^{(0)}_R(x_i;s,q,t)$,
and obtain their properties which can afterwards be used for their axiomatic definition,
independent of (\ref{EG}) and, if necessary, deviating from it.

The question is intimately related to $x\leftrightarrow y$ symmetry (spectral duality \cite{specdua}),
which is lost for $\xi(z)$ beyond the Macdonald locus and can be somehow restored by introducing
another arbitrary function $\tilde\xi(z)$ for the $y$-dependence,
satisfying a $\xi\leftrightarrow \tilde\xi$ duality,
which is, however, beyond the analysis in this text.

Here we construct a basis of symmetric functions in the following way: we define a symmetric function
$\Big[{\bf P^\xi}^{(k)}_R(x_i;s,q,t)\Big]_{symm}$ as a sum
\be\label{defsym}
\Big[{\bf P^\xi}^{(k)}_R(x_i;s,q,t)\Big]_{symm}:=\sum_{\mu\vdash |R|}
\left[{\bf P^\xi}^{(k)}_R(x_i;s,q,t)\right]_{\mu}{\rm Mon}_{_\mu}[\vec x]
\ee
of monomial symmetric polynomials  ${\rm Mon}_{_\mu}$,
\be
{\rm Mon}_{_\mu}[\vec x] = \sum_{\sigma\in S_N} x_{\sigma(i)}^{\mu_i},
\label{monpols}
\ee
where $\Big[\ldots\Big]_{\mu}$ denotes the coefficient in front of $\prod_ix_i^{\mu_i}$. 

The difference
${\bf P^\xi}^{(0)}_R(x_i;s,q,t)-\Big[{\bf P^\xi}^{(0)}_R(x_i;s,q,t)\Big]_{symm}$ is always vanishing for $N=2$ and
is non-vanishing for $N>2$ first time at level 4 for the single diagram $[3,1]$. For $N=4$ this example looks as
\be
{\bf P^\xi}^{(0)}_{[3,1]}(x_i;s,q,t)-\Big[{\bf P^\xi}^{(0)}_{[3,1]}(x_i;s,q,t)\Big]_{symm}
=F_0\left(x_1x_2x_3^2+(x_1x_2+x_1x_3+x_2x_3)x_4^2\right)\nn\\
F_0:=\zeta_2(1)^2-\zeta_2(1)\zeta_2(qt)-\zeta_2(1)\zeta_2(q)+\zeta_2(t)\zeta_2(qt)
\ee
where ${\rm Mon}_{R}^{(k)}$ denotes the monomial symmetric polynomial of variables $x_i$, $i=1,\ldots,k$.

Similarly, at level 5 one has:
\be
{\bf P^\xi}^{(0)}_{[4,1]}(x_i;s,q,t)-\Big[{\bf P^\xi}^{(0)}_{[4,1]}(x_i;s,q,t)\Big]_{symm}
=F_1\Big(x_1x_2x_3^3+(x_1x_2+x_1x_3+x_2x_3)x_4^3\Big)+\nn\\
+F_2\Big((x_1x_3^2+x_2x_3^2+x_2^2x_3+x_1^2x_2+x_2^2x_1+x_2^2x_3)x_4^2+x_1^2x_2x_3^2+x_1x_2^2x_3^2\Big)
+F_3x_1x_2x_3^2x_4+F_4x_1x_2x_3x_4^2
\ee
\be
F_1:=\zeta_2(1)\zeta_3(1)-\zeta_2(q^2t)\zeta_3(1)-\zeta_4(1)+\zeta_4(t)\nn\\
F_2:=\zeta_3(1)\Big(\zeta_3(1)-\zeta_2(1)\zeta_2(q^2t)-\zeta_3(q)+\zeta_3(qt)\Big)\nn\\
F_3:=\zeta_3(1)\Big(\zeta_2(1)^2\zeta_2(q^2t)-\zeta_2(q^2t)\zeta_3(1)-\zeta_2(1)\zeta_3(qt)+\zeta_3(qt)\Big)\nn\\
F_4:=\zeta_3(1)F_1
\ee
All these combinations $F_i$ are vanishing at the Macdonald locus.

\section{Role of the functor parameters}

\paragraph{The role of $N$.}

$N$ appears in sec.\ref{GenShirdef} in several roles:

\begin{itemize}
\item{As the number of variables $x_i$}
\item{In reduction condition for $y_i$}
\item{As the length of Young diagrams $R_i$}
\end{itemize}

In the leading order in $p$, the polynomial ${\bf P^\xi}^{(0)}_R(x_i;s,q,t)$ can be considered as a graded polynomial of time variables $p_k:=\sum_i^Nx_i^k$ so that $N$ can be considered arbitrary large, and $N$ does not enter formulas. In the usual terms of symmetric functions associated with representation theory, fixing concrete $N$ corresponds to reduction from $GL_\infty$ to $GL_N$.
A typical example is restricting the infinite Toda chain to a Toda chain of length $N$. This analogy can be pushed even further, see the next paragraph.

\paragraph{The role of $p$: series vs polynomials.}

The Shiraishi functions belong to the class of {\it mother functions},
i.e. they depend not on the Young diagrams, but on continuous variables $\vec y$.
Relation/reduction to Young diagram $R$ emerges through the specialization
\be
y_i = q^{R_i}(st)^{N-i}
\label{ydia}
\ee

The Shiraishi functions are
expressed only through $\vec x = \{x_1,\ldots,x_N\}$, no immediate lifting
to generic time-variables $p_k$ from the Miwa locus $p_k^* = \sum_{i=1}^N x_i^k$
is available at generic $\vec y$: they are {\it not} symmetric for generic $\vec y$.  Outside (\ref{ydia}), these functions are asymmetric series in $\vec x$,
interpolating between polynomials of various $R$-dependent degrees on
the loci (\ref{ydia}).

The Shiraishi functions depend on a variety of parameters.
From the point of view of the series-polynomial relation, distinguished is the role  of $p$.
The Shiraishi function is an infinite series, but it becomes a finite Laurent polynomial at each degree of $p$
after specializing $\vec y$ to a Young diagram by (\ref{ydia}).
Degree of $p$ is $p^{N\sum_{i,j=1}^N \lambda_{Nj-i+1}^{(i)}}$.
At the leading order $p^0$,  one gets after multiplication by $\prod_{i=1}^N x_i^{R_i}$ and symmetrization, an ordinary symmetric polynomial, which we call generalized Noumi-Shiraishi polynomial, since the mother function of this form for the Macdonald polynomials
was studied in detail in \cite{NS}.

The powers of $p$ are related to imposing the periodicity condition $x_{N+i}=x_i$. At a given $N$, $p$ dependence is expressed entirely through $p^N$, which can serve as a better parameter than $p$ itself. A counterpart of this parameter could be found in the periodic Toda chain: one can consider an infinite Toda chain and impose the periodicity condition with period $N$. Then, the Baker-Akhiezer function (the eigenfunction of the Lax operator) is quasi-periodic, and the counterpart of the quasiperiodicity parameter is just $p$. Bringing this parameter to zero, one obtains the open Toda chain, and this is much similar to the way of obtaining the generalized Noumi-Shiraishi polynomial. In algebraic terms, this corresponds to transition from the affine to finite-dimensional algebras.

One could also look for a description of the system at $N=\infty$ ($SL_\infty$ level), where $\vec x$ becomes
a sort of a function $x(p)$, and $p$ plays the role of the associated loop parameter.
Then one can pick up a particular locus (particular solutions of the would-be universal hypergeometric
equation), specified by a peculiar $N$-periodicity in order to get the Shiraishi series for particular $N$
with the condition $x_{N+i} = p^{-N}x_i$.

 {\bf The role of s} is still obscure to us, thus to avoid confusion we do not comment on it in the present text.

Note also that {\bf the role of parameters $q$ and $t$} becomes completely different in variance with the standard Macdonald case: $q$ governs the Pochhammer symbols and is in charge of the deformation from simple polynomials like the Schur polynomials, while $t$ is rather related to concrete details of this deformation. Moreover, $t$ in some parts of formulas is mixed with the parameter $s$ (like the choice of $y_i$), but in some other, not.

\section{GNS polynomials}

The polynomial ${\bf P^\xi}^{(0)}_R(x_i;s,q,t)$ for arbitrary function $\xi(z)$
does not depend on $s$, and so does the symmetric polynomial $\Big[{\bf P^\xi}^{(0)}_R(x_i;s,q,t)\Big]_{symm}$ thus we can denote it through
$\Big[{\bf P^\xi}^{(0)}_R(q,t)[x]\Big]_{symm}$.
It also does not depend on $N$ in the following sense:
it can be rewritten as a polynomial of the time variables
$p_k:=\sum_{i=1}^N x_i^k$,
and after that the coefficients of $\Big[{\bf P^\xi}^{(0)}_R(q,t)\{p_k\}\Big]_{symm}$
do not depend on $N$ (!).
Note that time variables $p_k$ are denoted by the same letter as the parameter $p$,
both notations are standard and can not be changed.
We hope that this will not cause confusion, especially because, in this paper,
we mostly discuss the GNS polynomials, which depend on $p_k$ but no longer on $p$
(they are the $p^0$ contribution to the Shiraishi series).

The GNS polynomial can be rewritten in the form, similar to the Noumi-Shiraishi representation
of the Macdonald polynomials \cite{NS}, hence, the name:
\be\label{GNS}
{\bf P^\xi}^{(0)}_R(q,t)[x]=\prod_{i=1}^N x_i^{R_i}
\cdot\sum_{m_{ij}}{\cal C}^R_n(m_{ij}|q,t)\prod_{1\le i<j\le N}\left({x_j\over x_i}\right)^{m_{ij}}
\ee
where $m_{ij}=0$ for $i\ge j$, $m_{ij}\in \mathbb{Z}_{\ge 0}$,

\be
\begin{array}{lc}
{\cal C}^R_n(m_{ij},|q,t):=&\\
= \prod_{k=2}^n\prod_{1\le i<j\le k}{\Xi\Big(q^{R_j-R_i+\sum_{a>k}(m_{ia}-m_{ja})}t^{i-j+1};q\Big)_{m_{ik}}
\over \Xi\Big(q^{R_j-R_i+\sum_{a>k}(m_{ia}-m_{ja})}qt^{i-j};q\Big)_{m_{ik}}}\cdot
\prod_{k=2}^n\prod_{1\le i\le j<k}{\Xi\Big(q^{R_j-R_i-m_{jk}+\sum_{a>k}(m_{ia}-m_{ja})}qt^{i-j-1};q\Big)_{m_{ik}}
\over \Xi\Big(q^{R_j-R_i-m_{jk}+\sum_{a>k}(m_{ia}-m_{ja})}t^{i-j};q\Big)_{m_{ik}}}&
\end{array}
\label{c}
\ee
Formulas (\ref{GNS}) and (\ref{defsym}) allow one to give a manifest representation of the GNS symmetric polynomials in terms of monomial symmetric polynomials  ${\rm Mon}_{_R}$,
\be
\Gns_R[x]:=\Big[{\bf P^\xi}^{(0)}_R(q,t)[x]\Big]_{symm}=\sum_{P}C_{RP}^{(q,t)} \cdot{\rm Mon}_{_P}[x]
\ee
Lifted to the time-variables space $p_k:=\sum_ix_i^k$, the monomial symmetric polynomials coincide with restriction of Macdonald polynomials
to $t=1$ (which do not depend on $q$):
\be
\begin{array}{rcl}
{\rm Mon}_{_R}[\vec x] &=& {\rm Mac}_{_R}^{t=1}[\vec x]\\
&\downarrow&\\
{\rm Mon}_{_R}\{p_k\}&=&{\rm Mac}_{_R}^{t=1}\{p_k\}
\end{array}
\ee
The Kostka matrix $C_{RP}^{(q,t)}$ is diagonal in the size of Young diagrams,
$C_{RP}\sim \delta_{|R|,|P|}$, and it is triangular in a matrix form, for the lexicographical ordering of Young diagrams:
\be\label{tr}
\boxed{
\Gns_R\{p_k\}={\rm Mon}_{_R}\{p_k\}+\sum_{P<R}C_{RP}^{(q,t)} \cdot{\rm Mon}_{_P}\{p_k\}
}
\ee
In the lexicographical ordering, $R>P$ if, for $i$ running from 1, the first non-vanishing difference $R_i-P_i$ is positive. 
In fact, similarly to the ordinary Macdonald polynomials, it is sufficient to choose the natural (or dominance) partial ordering \cite{Mac}, since the elements of the Kostka matrix between the unordered Young diagrams vanish. In this ordering, $R>P$ if $\sum_{i=1}^k\Big(R_i-P_i\Big)$ for all $k$. It is a partial ordering: for instance, the natural ordering does not fix the order of the diagrams [2,2,2] and [3,1,1,1], and $C_{[2,2,2],[3,1,1,1]}^{(q,t)}=C_{[3,1,1,1],[2,2,2]}^{(q,t)}=0$.

Thus, the non-deformed GNS polynomial $\Gns_{R}$ corresponds to the antisymmetric Young diagram, while the most deformed, to the symmetric one.

From representation (\ref{GNS}) it follows that the coefficients of GNS polynomials are graded combinations of products of the function $\eta(z):= \frac{\xi(qz)\xi(t/qz)}{\xi(tz)\xi(z)}$ at various points $q^it^j$. Hence, there is a hidden symmetry of the polynomials: they are invariant w.r.t. the replace $\xi(z)\to z^a\xi(z)$ with an arbitrary $a$.

If one chooses $\xi(z)=1-z$, then
$\Gns_R[x]$ is the standard Macdonald polynomial, and (\ref{GNS}) gives its Noumi-Shiraishi representation \cite{NS}.

\section{Properties of the GNS polynomials}

\paragraph{Orthogonality.} Let us define a conjugate system of polynomials in the following way. Denote $\Psi_{RQ}$ the coefficients of the $p$-expansion of the GNS
\be\label{ex1}
\Gns_{R}=\sum_Q \Psi_{RQ}\cdot p_Q
\ee
(this expansion is in no way triangular). Then, the set of polynomials
\be\label{ex2}
\Gns_{R}^\perp=\sum_Q \Psi^{-1}_{QR}\cdot{p_Q\over z_Q}
\ee
with $\Psi^{-1}$ being the inverse matrix, is orthogonal,
\be\label{or}
\Big<\Gns_{R}\Big|\Gns_{Q}^\perp\Big>=\delta_{RQ}
\ee
w.r.t. to the measure
\be\label{orm}
\left<p_\Delta\Big|p_{\Delta'}\right>\ =\ z_\Delta\delta_{\Delta,\Delta'}
\ee

\paragraph{Triangularity.}
The polynomials $\Gns_{R}^\perp$ possess a triangular expansion
\be\label{trD}
\Gns_{R}^\perp=\prod_i {\rm Schur}_{[R_i]}+\sum_{P>R}C_{RP}^\perp\prod_i {\rm Schur}_{[P_i]}
\ee
again with graded coefficients in $\eta$. This follows from the orthogonality relations
\be
\Big<{\rm Mon}_{R}\Big|\prod_i {\rm Schur}_{[Q_i]}\Big>=\delta_{RQ}
\ee
It also follows that $C_{RP}^{\perp}$ is just the transposition of the matrix inverse to $C_{RP}^{(q,t)}$, (\ref{tr}):
\be
C_{RP}^{\perp}=C^{-1}_{PR}
\ee
and the (partial) natural ordering is again sufficient in (\ref{trD}): the corresponding coefficients are zero like the previously discussed example: $C^\perp_{[2,2,2],[3,1,1,1]}=C^\perp_{[3,1,1,1],[2,2,2]}=0$.

Thus, the non-deformed $\Gns_{R}^\perp$ corresponds to the symmetric Young diagram, while the most deformed, to the antisymmetric one.

Note that, in the Macdonald case,
\be
\Gns_{R}^\perp\Big|_{\xi(z)=1-z}\{p_k\}={\rm\overline{Mac}}_{R^\vee}\{(-1)^{k+1}p_k\}
\ee
where the bar over the Macdonald polynomial denotes the permutation of $q$ and $t$.

\paragraph{The Cauchy formula.}

Having the orthogonality relation (\ref{or}), one can also immediately get the Cauchy formula for the GNS polynomials:
\be\label{Cauchy}
\sum_R \Gns_{R}\{p_k\}\cdot \Gns_{R}^\perp\{p_k'\}=\exp\left(\sum_k{p_kp_k'\over k}\right)
\ee
It follows from (\ref{ex1}) and (\ref{ex2}):
\be
\sum_R \Gns_{R}\{p_k\}\cdot \Gns_{R}^\perp\{p_k'\}=
\sum_{R,Q,Q'} \Psi_{RQ}\cdot \Psi^{-1}_{Q'R}\cdot{p_Qp'_{Q'}\over z_{Q'}}
=\sum_{Q} {p_Qp'_{Q}\over z_{Q}}=
\exp\left(\sum_k{p_kp_k'\over k}\right)
\ee

\paragraph{Ring structure.}
One can construct the structure constants of the ring
formed by the polynomials ${\bf P^\xi}^{(0)}_R(q,t)$,
i.e. the generalized Littlewood-Richardson coefficients:
\be
\Gns_{R'}\{p_k\}\cdot\Gns_{R''}\{p_k\}=
\sum_R{\bf N}^R_{R'R''}\Gns_R\{p_k\}
\ee
They possess an interesting property: in the case of Kerov functions,
non-vanishing are only the coefficients between the partitions
$R'\cup R'' = [r'_1+r''_1,r'_2+r_2'',\ldots]$ and
$R'+R'' = [{\rm ordered\ collection\ of\ all}\ r'_i \ {\rm and} \ r''_j]$;
in the case of Macdonald polynomials, there are additional zeroes,
since only the partitions associated with the irreducible representations
(via the Schur-Weyl duality)
emerging in the decomposition of $R'\otimes R''$ contribute, and not all of
the representations between $R'\cup R''$ and $R'+R''$
(in the partial ordering) emerge in this decomposition.
In the GNS case, similarly to the Macdonald case,
the coefficients ${\bf N}^R_{R'R''}$ are vanishing
for the partitions with the ordering number larger than $R'\cup R''$,
and vanishing for the representations between $R'\cup R''$ and
$R'+R''$ for $R$ not belonging to the decomposition of $R'\otimes R''$.
However, they are non-vanishing for partitions with the ordering number less than $R'+R''$.
For instance, the group theory decomposition dictates that, for the Macdonald polynomials,
\be
{\rm Mac}_{[2]}\cdot {\rm Mac}_{[1,1]}={\rm Mac}_{[3,1]}\oplus {\rm Mac}_{[2,1,1]}
\ee
and, in the more general case of Kerov functions,
\be
{\rm Kerov}_{[2]}\cdot {\rm Kerov}_{[1,1]}
={\rm Kerov}_{[3,1]}\oplus {\rm Kerov}_{[2,2]}\oplus {\rm Kerov}_{[2,1,1]}
\ee
At the same time, for the GNS,
\be
\Gns_{[2]}\cdot \Gns_{[1,1]}=
\Gns_{[3,1]}\oplus \Gns_{[2,1,1]}
\oplus \Gns_{[1,1,1,1]}
\ee

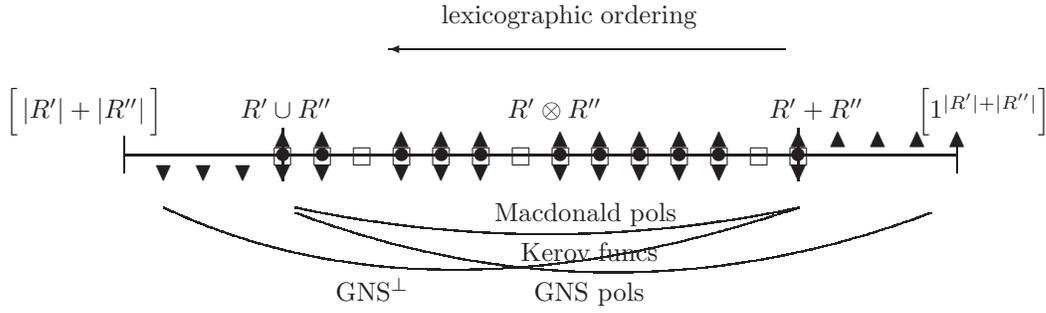
\begin{figure}
\begin{picture}(200,100)(-70,-50)
\put(5,0){\line(1,0){315}}
\put(-40,14){\mbox{$\Big[\,|R'|+|R''|\,\Big]$}} \put(5,-7){\line(0,1){14}}
\put(305,14){\mbox{$\left[1^{|R'|+|R''|}\right]$}} \put(320,-7){\line(0,1){14}}
\put(49,14){\mbox{$R'\cup R''$}} \put(65,-10){\line(0,1){20}}
\put(249,14){\mbox{$R'+R''$}} \put(260,-10){\line(0,1){20}}
\put(150,14){\mbox{$R'\otimes R''$}}
\put(65,0){\circle*{5}}
\put(61.1,-4){$\square$}
\put(76.1,-4){$\square$}
\put(106.1,-4){$\square$}
\put(121.1,-4){$\square$}
\put(136.1,-4){$\square$}
\put(166.1,-4){$\square$}
\put(181.1,-4){$\square$}
\put(196.1,-4){$\square$}
\put(211.1,-4){$\square$}
\put(226.1,-4){$\square$}
\put(256.1,-4){$\square$}
\put(271.3,3){$\blacktriangle$}
\put(286.3,3){$\blacktriangle$}
\put(301.3,3){$\blacktriangle$}
\put(316.3,3){$\blacktriangle$}
\put(61.3,3){{$\blacktriangle$}}
\put(76.3,3){{$\blacktriangle$}}
\put(91.1,-4){{$\square$}}
\put(106.3,3){{$\blacktriangle$}}
\put(121.3,3){{$\blacktriangle$}}
\put(136.3,3){{$\blacktriangle$}}
\put(166.3,3){{$\blacktriangle$}}
\put(181.3,3){{$\blacktriangle$}}
\put(196.3,3){{$\blacktriangle$}}
\put(211.3,3){{$\blacktriangle$}}
\put(226.3,3){{$\blacktriangle$}}
\put(241.1,-4){{$\square$}}
\put(256.3,3){{$\blacktriangle$}}
\put(16.3,-9.8){{$\blacktriangledown$}}
\put(31.3,-9.8){{$\blacktriangledown$}}
\put(46.3,-9.8){{$\blacktriangledown$}}
\put(61.3,-9.8){{$\blacktriangledown$}}
\put(76.3,-9.8){{$\blacktriangledown$}}
\put(106.3,-9.8){{$\blacktriangledown$}}
\put(121.3,-9.8){{$\blacktriangledown$}}
\put(136.3,-9.8){{$\blacktriangledown$}}
\put(151.1,-4){{$\square$}}
\put(166.3,-9.8){{$\blacktriangledown$}}
\put(181.3,-9.8){{$\blacktriangledown$}}
\put(196.3,-9.8){{$\blacktriangledown$}}
\put(211.3,-9.8){{$\blacktriangledown$}}
\put(226.3,-9.8){{$\blacktriangledown$}}
\put(256.3,-9.8){{$\blacktriangledown$}}
\put(80,0){\circle*{5}} \put(110,0){\circle*{5}}\put(125,0){\circle*{5}}
\put(140,0){\circle*{5}}\put(170,0){\circle*{5}}
\put(185,0){\circle*{5}} \put(200,0){\circle*{5}} \put(215,0){\circle*{5}}\put(230,0){\circle*{5}}
\put(260,0){\circle*{5}}
\put(145,-25){\mbox{Macdonald pols}}
\qbezier(70,-20)(165,-40)(260,-20)
\put(155,-40){\mbox{Kerov funcs}}
\qbezier(70,-22)(190,-67)(310,-22)
\put(160,-55){\mbox{GNS pols}}
\qbezier(20,-20)(120,-67)(260,-20)
\put(85,-55){\mbox{GNS$^\perp$}}
\put(255,40){\vector(-1,0){150}}
\put(125,50){\mbox{lexicographic ordering}}
\end{picture}
\caption{{\footnotesize
Symbolic description of zeroes of the product decomposition
$P_{R'}P_{R''}=\sum_Q P_Q$ in the lexicographically ordered
set of partitions (Young diagrams)  $Q$.
A product of Kerov functions ($\square$) gets non-vanishing contributions
from everywhere in between $R'\cup R''$ and $R'+R''$.
Contributions to the product of Macdonald (and Schur) polynomials ($\bullet$)
are non-vanishing only at {\it some} points in this domain,
which belong to the product of representations $R'\otimes R''$.
A product of the generalized Noumi-Shiraishi polynomials ($\blacktriangle$) can get non-vanishing contributions from
the same points as Macdonald polynomials, but also from the region in between
$R'+R''$ and $[1^{|R'|+|R''|}]$, while their conjugate polynomials ($\blacktriangledown$) gets contributions from
the same points as Macdonald polynomials, but also from the region in between
$R'\cup R''$ and $[|R'|+|R''|-1,1]$.
}}
\end{figure}

One can similarly define the generalized Littlewood-Richardson coefficients for the conjugate GNS:
\be
\Gns^\perp_{R'}\{p_k\}\cdot\Gns^\perp_{R''}\{p_k\}=
\sum_R{\bf N^\perp}^R_{R'R''}\Gns^\perp_R\{p_k\}
\ee
One may think these polynomials behave exactly like the Macdonald polynomials: the non-vanishing ring structure coefficients are determined by the group theory decomposition. For instance,
\be
\Gns^\perp_{[2]}\cdot \Gns^\perp_{[1,1]}=\Gns^\perp_{[3,1]}\oplus \Gns^\perp_{[2,1,1]}
\ee
However, this is not always the case: for instance,
\be
\Gns^\perp_{[1,1]}\cdot \Gns^\perp_{[1,1]}=\Gns^\perp_{[3,1]}\oplus
\Gns^\perp_{[2,2]}\oplus \Gns^\perp_{[2,1,1]}\oplus \Gns^\perp_{[1,1,1,1]}
\ee
while
\be
{\rm Mac}_{[1,1]}\cdot {\rm Mac}_{[1,1]}={\rm Mac}_{[2,2]}\oplus {\rm Mac}_{[2,1,1]}\oplus {\rm Mac}_{[1,1,1,1]}
\ee
Generally, the coefficients ${\bf N^\perp}^R_{R'R''}$ are vanishing
for the partitions with the ordering number less than $R'\cup R''$,
and vanishing for the partitions between $R'\cup R''$ and
$R'+R''$ for $R$ not belonging to the decomposition of representations $R'\otimes R''$.
However, they are non-vanishing for partitions with the ordering number larger than $R'+R''$ for exception of the partition
associated with the symmetric representation $[|R'|+|R''|]$.

\paragraph{Skew polynomials.}

The skew GNS polynomials can be defined in the standard way:
\be
\Gns_R\{p_k+p_k'\}=\sum_Q\Gns_{R/Q}\{p_k\}\cdot\Gns_Q\{p_k'\}
\ee
and one can similarly define the dual skew polynomials $\Gns^\perp_{R/Q}\{p_k\}$.
Properties of the skew polynomials follow from the orthogonality relations. In particular, similarly to the standard Macdonald (and further, the Kerov case), the skew GNS polynomials can be constructed from the ring structure coefficients:
\be\label{skewN}
\Gns_{R/Q}=\sum_P{{\bf N}^\perp}^R_{QP}\cdot\Gns_{P}
\ee
This follows from the Cauchy formula (\ref{Cauchy}) and the general theorem \cite{Cauchyskew}:

\be
\begin{array}{ccc}
\sum_R \Gns_R^\perp\{p\}\cdot \Gns_R \{p'+p''\}
&= \exp\left(\sum_k \frac{p_k(p'_k+p''_k)}{k}\right)
&= \sum_{P,Q} \Gns_P^\perp\{p\}\cdot \Gns_P\{p' \} \Gns_Q^\perp\{p\}\cdot \Gns_Q\{p'' \}\\
&&\\
|| && || \\
&&\\
\sum_{R,Q} \Gns_R^\perp\{p\}\cdot\Gns_{R/Q}\{p'\}\Gns_Q\{p''\}
&=& \sum_{P,Q,R} {{\bf N}^\perp}_{PQ}^R \Gns_R^\perp\{p\}\Gns_P\{p' \}\Gns_Q\{p'' \}
\end{array}
\nn
\ee

\noindent
Assuming that the map $F:\ f(R,Q,p')\mapsto \tilde f(p,p',p"):=\sum_{R,Q}\Gns_R^\perp\{p\}\Gns_Q\{p'' \}f(R,Q,p')$ does not have a kernel, one can omit the identical factors $\Gns_R^\perp\{p\}\Gns_Q\{p'' \}$ at the two sides
of the bottom line, and get (\ref{skewN}).

One can also define the conjugate skew symmetric polynomials
\be
\Gns^\perp_R\{p_k+p_k'\}=\sum_Q\Gns^\perp_{R/Q}\{p_k\}\cdot\Gns^\perp_Q\{p_k'\}
\ee
and they are also given by the formula involving the ring structure coefficients:
\be\label{skewN2}
\Gns^\perp_{R/Q}=\sum_P{{\bf N}}^R_{QP}\cdot\Gns^\perp_{P}
\ee

\paragraph{The generalized Cauchy formula.}

As usual, one can easily extend the Cauchy formula (\ref{Cauchy}) to the case of skew polynomials:
\be\label{gCauchy}
\sum_R \Gns_{R/\eta_1}\{p_k\}\cdot \Gns_{R/\eta_2}^\perp\{p_k'\}=\exp\left(\sum_k{p_kp_k'\over k}\right)
\sum_\rho \Gns_{\eta_2/\rho}\{p_k\}\cdot \Gns_{\eta_1/\rho}^\perp\{p_k'\}
\ee
It is proved much similar to the previous formulas: using the map $F$. Indeed, apply this map to the l.h.s. of (\ref{gCauchy}) and use the definition of the skew GNS polynomials, formulas (\ref{skewN})-(\ref{skewN2}) and the Cauchy formula (\ref{Cauchy}):
\be
\sum_{\eta_1,\eta_2}\Gns_{\eta_1}\{\bar p_k\}\Gns^\perp_{\eta_2}\{\bar p_k'\}
\sum_R \Gns_{R/\eta_1}\{p_k\}\cdot \Gns_{R/\eta_2}^\perp\{p_k'\}=\sum_R
\Gns_{R}\{p_k+\bar p_k\}\cdot \Gns_{R}^\perp\{p_k'+\bar p'_k\}=\nn\\
=\exp\left(\sum_k{(p_k+\bar p_k)(p_k'+\bar p'_k)\over k}\right)=
\exp\left(\sum_k{p_kp_k'\over k}\right)\cdot \exp\left(\sum_k{p_k\bar p_k'\over k}\right)\cdot
\exp\left(\sum_k{\bar p_kp_k'\over k}\right)\cdot \exp\left(\sum_k{\bar p_k\bar p_k'\over k}\right)=\nn\\
=\exp\left(\sum_k{p_kp_k'\over k}\right)\cdot\sum_{\lambda,\mu,\rho}
\Gns_{\lambda}\{p_k\}\cdot \Gns_{\lambda}^\perp\{\bar p'_k\}\cdot
\Gns_{\mu}\{\bar p_k\}\cdot \Gns_{\mu}^\perp\{p'_k\}\cdot
\Gns_{\rho}\{\bar p_k\}\cdot \Gns_{\rho}^\perp\{\bar p'_k\}=\nn\\
=\exp\left(\sum_k{p_kp_k'\over k}\right)\cdot
\sum_{\lambda,\mu,\rho,\eta_1,\eta_2} \Gns_{\lambda}\{p_k\} \Gns_{\mu}^\perp\{p'_k\}\cdot
{{\bf N}^\perp}^{\eta_2}_{\lambda\rho}\Gns_{\eta_2}\{\bar p'_k\}\cdot
{{\bf N}}^{\eta_1}_{\mu\rho}\Gns_{\eta_1}\{\bar p_k\}=\nn\\
=\exp\left(\sum_k{p_kp_k'\over k}\right)\cdot
\sum_{\eta_1,\eta_2}\Gns_{\eta_1}\{\bar p_k\}\Gns^\perp_{\eta_2}\{\bar p_k'\}\cdot
\sum_\rho \Gns_{\eta_2/\rho}\{p_k\}\cdot \Gns_{\eta_1/\rho}^\perp\{p_k'\}\nn
\ee
Then, again, formula (\ref{gCauchy}) follows from the assumption of absence of a kernel of the map $F$.

\paragraph{The Hamiltonian.}

One of the main features of the standard symmetric functions is that they satisfy the eigenvalue equation with a Hamiltonian \cite{Rui,MMHam}. In fact, there are typically infinitely many Hamiltonians, which is a consequence of integrability behind a set of symmetric functions, but the first Hamiltonian is sufficient to unambiguously restore the whole set. Hence, one has to look for a Hamiltonian for the GNS polynomials. At $N=2$ case, they satisfy the equation
\be
\hat H{\rm Gns}_{[r]}=0,\nn\\
\hat H:={x_1\xi(\hat P_2)\xi(t\hat P_1)-x_2\xi(\hat P_1)\xi(t\hat P_2)\over x_1-x_2}
\ee
where $P_i:=q^{x_i\partial_{x_i}}$. Moreover,
\be
\hat H{\rm Gns}_{[r,s]}=\xi(q^s)\xi(tq^r)\cdot{\rm Gns}_{[r,s]}+\sum_{1\le i\le (r-s)/2}\alpha_i\cdot {\rm Gns}_{[r-i,s+i]}
\ee
and the Young diagrams with more than 2 rows do not contribute in the $N=2$ case. This means that ${\rm Gns}_{[r,s]}$ with $s=r,r-1$ are also eigenfunctions along with ${\rm Gns}_{[r]}$, i.e. with $s=0$, the eigenvalues being $\xi(q^s)\xi(tq^r)$.

Let us remind that the standard Macdonald Hamiltonian has the form
\be
\hat H={x_2-tx_1\over x_2-x_1}\hat P_1+{x_1-tx_2\over x_1-x_2}\hat P_2
\ee

\bigskip

To illustrate how it works, consider the action of this Hamiltonian on the $\Gns_{[2]}$:
\be
\xi(\hat P_2)\xi(t\hat P_1)x_1^2=x_1^2\xi(1)\xi(tq^2)=0,\ \ \ \ \ \ \xi(\hat P_1)\xi(t\hat P_2)x_1^2=x_1^2\xi(q^2)\xi(t)\nn\\
\xi(\hat P_2)\xi(t\hat P_1)x_1x_2=x_1x_2\xi(q)\xi(tq),\ \ \ \ \ \ \ \xi(\hat P_1)\xi(t\hat P_2)x_1x_2=x_1x_2\xi(q)\xi(qt)\nn \\
\xi(\hat P_2)\xi(t\hat P_1)x_2^2=\xi(q^2)\xi(t),\ \ \ \ \ \ \ \xi(\hat P_1)\xi(t\hat P_2)x_2^2=x_2^2\xi(tq^2)\xi(1)=0
\ee
Since
\be
\Gns_{[2]}=x_1^2+x_2^2+{\xi(q^2)\xi(t)\over\xi(q)\xi(qt)}x_1x_2
\ee
one immediately obtains
\be
\Big(x_1\xi(\hat P_2)\xi(t\hat P_1)-x_2\xi(\hat P_1)\xi(t\hat P_2)\Big)\Gns_{[2]}=0
\ee
It turns out to be a general rule at any $N$: the GNS polynomial associated with one-row partition satisfies the equation
\be
\hat H{\rm Gns}_{[r]}=0,\nn\\
\hat H:=\prod_{i<j} \Big(x_i\xi(\hat P_j)\xi(t\hat P_i)-x_j\xi(\hat P_i)\xi(t\hat P_j)\Big)
\ee
The standard Macdonald Hamiltonian for arbitrary $N$ has the form
\be\label{MHam}
\hat H=\sum_i\prod_{j\ne i}{x_j-tx_i\over x_j-x_i}\hat P_i
\ee

Note that this Hamiltonian depends explicitly on the original function $\xi(z)$,
while their eigenfunctions depend on $\eta(z)$.
In certain sense, the Hamiltonian has an ``anomaly" as compared to the eigenfunctions
(symmetric polynomials $\Gns_R$).

Since $\Gns^\perp$ associated with one-row partition is just the non-reformed Schur polynomial, it is trivially the eigenfunction of the usual Macdonald Hamiltonian (\ref{MHam}).

Thus, we demonstrated that the GNS are associated, at least, with quasi-exactly solvable system \cite{TU}.

\section{Incompatibility with Kerov functions}

In this section, we  explain that the GNS polynomials are an extension of the Macdonald polynomials very different from another extension known under the name Kerov functions. We deal with the very explicit polynomials of the first levels, the formulas for the GNS case being collected in the Appendix.

\bigskip

{\bf 1.}
Kerov functions were introduced in \cite{Kerov} and recently reviewed in some detail
in \cite{MMkerov}.
They are obtained by a {\it triangular} orthogonalization of polynomials
(\ref{monpols})
w.r.t. the scalar product
\be
\left< p_{R}|p_{R'}\right> = \delta_{R,R'}\cdot z_R\cdot g_R
\ee
$g_R:=\prod_ig_{R_i}$ with arbitrary parameters $g_k$.
The triangular Kostka matrix ${\rm Kerov}_R = \sum_Q C_{RQ} \cdot {\rm Mon}_{_Q}$
for the Kerov functions looks as follows:
\be
C^{(1)} = \left(1\right) \nn \\  \nn \\
C^{(2)} = \left(\begin{array}{cc} 1 & 0 \\ \\ \frac{2g_2}{g_1^2+g_2} & 1 \end{array}\right)
\nn \\ \nn \\ \nn \\
C^{(3)} = \left(\begin{array}{ccc} 1 & 0
& 0   \\ \\
\frac{3g_3(g_1^2+g_2) }{\Delta_3^\vee} & 1 &   0 \\ \\
\frac{6g_2g_3}{\Delta_3^\vee} & \frac{6(g_1g_2+g_3)}{\Delta_3} & 1 \end{array}\right)
\nn
\ee

\bigskip

$$
C^{(4)} =
$$

{\footnotesize
$$
 =\left(\begin{array}{ccccc}
1 &0
&0
& 0
&0
\\ \\
\frac{4g_2g_4(2g_1^3g_2+3g_1^2g_3+g_2g_3)}{\Delta_4^\vee} & 1 & 0
& 0
& 0
\\ \\
 \frac{6g_3g_4(g_2^2+g_2)^2}{\Delta_4^\vee} & \!\!\!\!\!\!\!\!\!\!\!\!\!\!\!\!\!\!\!\!\!\!\!\!
-\frac{2g_2(g_1^2+g_2)(g_1^3g_4-3g_1^2g_2g_3-3g_1g_2g_4-g_3g_4)}{\Delta_4'^\vee} & 1 & 0
& 0
\\ \\
\frac{12g_2g_3g_4(g_1^2+g_2)}{\Delta_4^\vee} & \frac{\begin{array}{cc}6g_1^4g_2^2g_3+3g_1^4g_3g_4+4g_1^3g_2^2g_4+\\
+12g_1^2g_2^3g_3 +6g_1^2g_2g_3g_4+12g_1g_2^3g_4+5g_2^2g_3g_4\end{array}}
{\Delta_4'^\vee} & \frac{2(g_2^2+g_4)(g_1^3+3g_1g_2+2g_3)}{\Delta_4'} &  1 & 0 
\\ \\
 \frac{24g_2^2g_3g_4}{\Delta_4^\vee} & \frac{12g_2(2g_1^2g_2^2g_3+g_1^2g_3g_4+2g_1g_2^2g_4+g_2g_3g_4)}{\Delta_4'^\vee} & \frac{12(g_2^2+g_4)(g_1g_2+g_3)}{\Delta_4'} & \frac{12(g_1^2g_2+2g_1g_3+g_2^2+2g_4)}{\Delta_4} & 1
\end{array}\right)
$$
}

\bigskip

\be
\ldots
\nn \\ \nn \\
\boxed{
C_{[3,1,1,1],[2,2,2]}\ \stackrel{{\footnotesize \hbox{\cite[eq.(37)]{MMkerov}}}}{\sim} \
x_1^2x_2^2x_3^2(x_1^2-x_2x_3)(x_2^2-x_1x_3)(x_3^2-x_1x_2)\cdot {\rm Schur}_{[7]}\{g\}\neq 0
}
\nn \\ \nn \\
\ldots
\label{KerKost}
\ee
with denominator functions $\Delta$ described in \cite[eq.(71)]{MMkerov}:
\be
\Delta_3 = g_1^3+3g_1g_2+2g_3 = 3!{\rm Schur}_{[3]}\{g\} \nn \\
\Delta_3^\vee = 2g_1^3g_2+3g_1^2g_3+g_2g_3 = 3!g_1^2g_2g_3{\rm Schur}_{[3]}\{g^{-1}\} \nn \\
\Delta_4^\vee = 6g_1^4g_2^2g_3+3g_1^4g_3g_4+8g_1^3g_2^2g_4+6g_1^2g_2g_3g_4+g_2^2g_3g_4\nn\\
\Delta_4'^\vee = 2g_1^5g_2^3+g_1^5g_2g_4+6g_1^4g_2^2g_3+3g_1^4g_3g_4+2g_1^3g_2^2g_4
+4g_1^2g_2^3g_3+2g_1^2g_2g_3g_4+3g_1g_2^3g_4+g_2^2g_3g_4
\ee

\bigskip

{\bf 2.}
There is a minor degree of consistency between (\ref{KerKost}) and (\ref{GNSC}),
for example, in (\ref{GNSC})  $C_{[3][111]}=C_{[3][21]}C_{[2][11]}$, and the same is true
for (\ref{KerKost}):
\be
\frac{6g_2g_3}{\Delta_3^\vee} = \frac{3g_3(g_1^2+g_2)}{\Delta_3^\vee} \cdot \frac{2g_2}{g_1^2+g_2}
\ee
or, say, $C_{[4][22]}=C_{[4][31]}C_{[3][21]}$
\be
\frac{6g_3g_4(g_2^2+g_2)^2}{\Delta_4^\vee} =
\frac{4g_2g_4(2g_1^3g_2+3g_1^2g_3+g_2g_3)}{\Delta_4^\vee}
\cdot \frac{3g_3(g_1^2+g_2) }{\Delta_3^\vee}
\ee
These relations demonstrate that sometime the numerators in (\ref{KerKost})
are made from the same denominator functions, but generically these relations
are far more involved.
The same is true about the consistency domain between (\ref{KerKost}) and (\ref{GNSC}): it is too small.
Despite the ``number of degrees of freedom" in the both cases is the same
(it is given by an arbitrary function, or by an infinite set of coefficients), they enter the Kostka matrices
in a very different way, and the
elements of (\ref{KerKost}) are severely stronger constrained than the function $\xi$.

\bigskip

{\bf 3.}
There is no room for the {\it second} set of  GNS polynomials,
as it happens in the Kerov case.
Normally the second set of polynomials is associated with a transformation
like (\ref{KerKost}), which is triangular w.r.t. the alternative, anti-lexicographic
ordering of Young diagrams.
However, in the GNS case there is no difference:
the second set of polynomials is just the same, the change of ordering does not change
the polynomials.
This follows from the vanishing of the Kostka matrix elements between partitions differing in
the two different orderings. The first such a matrix element emerges at level 6 between the partitions
$[2,2,2]$ and $[3,1,1,1]$: $C_{[2,2,2],[3,1,1,1]}=C_{[3,1,1,1],[2,2,2]}=0$ (and similarly for their transposed
$[3,3]$ and $[4,1,1]$).

Moreover, it is shown in \cite{MMkerov,MMcomp} that, within Kerov family, these matrix elements
vanish only for the Macdonald polynomials, therefore, one can conclude that the intersection
\be
{\rm GNS\ pols} \  \cap  \ {\rm Kerov\ funcs} \  =\  {\rm Macdonald\ pols}
\ee

{\bf 4.}
Orthogonality of the GNS polynomials implies
non-diagonality of the scalar product $\left<p_{R}\Big| p_{R'}\right>$.
For example, at level $3$, the diagonalizability of
$C^{(3)}=\left(\begin{array}{ccc} 1 & 0& 0 \\ a& 1& 0\\ b& c& 1\end{array}\right)$
requires that
$c = \frac{(a-3)b}{a^2-a-b}$.
It is indeed true for the Kerov case, when
\be
a =  \frac{3g_3(g_1^2+g_2) }{\Delta_3^\vee},
\nn \\
b = \frac{6g_2g_3}{\Delta_3^\vee},
\nn \\
c = \frac{6(g_1g_2+g_3)}{\Delta_3}
\ee
but is not true for (\ref{GNSC}).

In fact, even a little bit more strong statement is correct: the
requirement of diagonal scalar product of the GNS polynomials is
inconsistent with a diagonal scalar product of the Kerov functions.

\section{Conclusion}

In this paper, we introduced a new infinite-parametric family of symmetric polynomials,
which seems to be an interesting generalization of the two-parametric Macdonald polynomials.
They are also obtained by a triangular transformation of monomial symmetric polynomials and Schur polynomials;
they form a closed ring under multiplication graded by the size of the Young diagrams;
they are independent on the choice of ordering of Young diagrams within the natural partial ordering;
and their ring structure coefficients lying between $R'\cup R''$ and
$R'+R''$ vanish together with those in the Macdonald case,
i.e. for representations which do not appear in the product of other two:
\be
R_1+R_2\le Q\le R_1\cup R_2:\ \ \ \
{{\bf N}^\perp}^Q_{R_1R_2} \neq 0 \ \ \Longleftrightarrow \ \ Q\in R_1\otimes R_2,
\label{gtprop}
\ee
(note that there is still a deviation from representation theory:
some ${\bf N}^\perp$ have ``tails", they do not vanish also for some ``more symmetric" $Q>R_1+ R_2$,
while ${\bf N}$, though also satisfying (\ref{gtprop}),
have tails from the other, ``antisymmetric" side: for $Q<R_1\cup R_2$).

In addition, the Kostka matrix relating GNS polynomials to the Schur ones has a non-trivial grading.
This makes this family an interesting alternative, or, better, complement,
to the family of Kerov functions \cite{Kerov,MMkerov}, which is equally large,
but deviates much more from representation theory.

This remarkable family is currently built in a four-step process.

\begin{itemize}

\item{ First}, we consider a far-going generalization of Shiraishi series \cite{S},
by introducing an arbitrary function $\xi(z)$, instead of just $\xi(z) = 1-z$
in \cite{S} or $\xi(z)\sim\vartheta(z)$ in \cite{FOS,AKMM2} related to the Dell
deformation of the integrable Calogero-Ruijsenaars systems and Nekrasov theories.
We call this generalization {\it Shiraishi functor} from the space of functions $\xi(z)$
restricted only by behaviour w.r.t. $z\to 1/z$
to that of hypergeometric-like Shiraishi series.

\item{Second}, we restrict $\vec y$ variables in Shiraishi {\it mother function}
to Young diagrams by the usual rule (\ref{ydia}),
expand in peculiar parameter $p$ and, after an appropriate rescaling, pick up a $p^0$ term.
This procedure mimics that in \cite{NS}, and we call the result
the {\it Generalized Noumi-Shiraishi} symmetric {\it polynomials} in $\vec x$.

\item{Third}, we expand these polynomials in, say, Schur functions
(expansion appears to be triangular),
what allows to raise them from the space of $\vec x$ to that of time-variables $p_k$.
This provides what we call GNS polynomials, $\Gns_R^\xi\{p\}$.

\item{Forth}, we construct a system of polynomials $\boxed{\Gns^\perp_R\{p\}}$,
conjugate to $\Gns_R\{p\}$
w.r.t. the standard scalar product of time-variables
$\left< p_\Delta\Big| p_{\Delta'}\right>\,=\,z_\Delta \delta_{\Delta,\Delta'}$.
These polynomials appear to be remarkably distinguished by their properties.

\end{itemize}

This system possesses further generalization by the change of the scalar product
and by a look on the higher powers in $p$.
Calculations in the latter case become far more tedious and will be considered elsewhere.
The primary task for the future research is to further explore these ,$\Gns_R\{p\}$, $\Gns^\perp_R\{p\}$,
with the hope to find their straightforward definition,
for example, by finding a $\xi$-dependent scalar product, in which they become
self-orthogonal (rather than form a {\it bi-orthogonal} system with the GNS polynomials $\Gns_R\{p\}$),
or by finding a general formula for the $\xi$-deformation of the Ruijsenaars
Hamiltonians, for which $\Gns^\perp_R\{p\}$ are the common eigenfunctions.

Clearly, the remarkable polynomials $\Gns^\perp_R\{p\}$, or their symmetric-polynomial
version $\Gns^\perp_R[x]$ on the Miwa locus $p_k = \sum_{i=1}^N x_i^k$,
deserve a separate name and notation, free of the reference to scaffolding used in our
current construction .
This is, however, a little premature to decide, first the adequate definitions
and language should be found.

\section*{Acknowledgements}

We appreciate useful discussions with   Y.Zenkevich.

Our work is supported in part by
Grants-in-Aid for Scientific Research (17K05275) (H.A.), (15H05738, 18K03274) (H.K.)
and JSPS Bilateral Joint Projects (JSPS-RFBR collaboration)
``Elliptic algebras, vertex operators and link invariants'' from MEXT, Japan.
It is also partly supported by the grant of the
Foundation for the Advancement of Theoretical Physics ``BASIS" (A.Mir., A.Mor.),
by  RFBR grants 19-01-00680 (A.Mir.) and 19-02-00815 (A.Mor.),
by joint grants 19-51-53014-GFEN-a (A.Mir., A.Mor.), 19-51-50008-YaF-a (A.Mir.),
18-51-05015-Arm-a (A.Mir., A.Mor.), 18-51-45010-IND-a (A.Mir., A.Mor.).
The work was also partly funded by RFBR and NSFB according
to the research project 19-51-18006 (A.Mir., A.Mor.).

\section*{Appendix. Explicit examples of GNS and GNS$^\perp$}

Hereafter, we introduce the notation
\be
\zeta_k(z) =   \frac{\xi(q^kz)\xi(tz)}{\xi(q^{k-1}tz)\xi(qz)} = \prod_{i=1}^{k-1} \eta(q^iz)
\ee

\subsection*{Triangular expansion of the GNS polynomials}

The triangular expansion of the GNS polynomials is given by (\ref{tr}),
$\Gns_R\{p\} = \sum_{Q} C_{RQ} \cdot {\rm Mon}_Q\{p\}$ with
\be\label{GNSC}
C^{(1)} = (1)
\nn \\ \nn \\
C^{(2)} = \left(\begin{array}{c||ccc} & & [1,1] & 2 \\ \hline \\
\phantom.[1,1] & & 1 & 0 \\  \phantom.[2] & & \zeta_2(1) & 1
\end{array}\right)
\nn \\ \nn \\ \nn \\
C^{(3)} = \left(\begin{array}{ccc} 1 & 0 & 0 \\ \\
\zeta_2(1) +\zeta_2(t) & 1 & 0 \\ \\
\zeta_2(1)\zeta_3(1) & \zeta_3(1) & 1 \\ \\
\end{array}\right)
\nn \\ \nn \\ \nn \\
C^{(4)} = \left(\begin{array}{ccccc}
1 & 0 & 0 & 0 & 0 \\ \\
\zeta_2(1)+\zeta_2(t)+\zeta_2(t^2) & 1 & 0 & 0 & 0 \\ \\
\zeta_2(1)\Big(\zeta_2(1)+\zeta_2(t)\Big) &   \zeta_2(1)^2\zeta_2(qt) & 1 & 0 & 0 \\ \\
\zeta_2(1)\Big(\zeta_2(1)\zeta_2(q)+\zeta_2(1)\zeta_2(qt)+\zeta_2(t)\zeta_2(qt)\Big)
&\zeta_2(1)\Big(\zeta_2(q)+\zeta_2(qt)\Big)& \zeta_2(1)   &1 & 0 \\ \\
\zeta_2(1)\zeta_3(1)\zeta_4(1) & \zeta_3(1)\zeta_4(1) &  \zeta_3(1)\zeta_3(q) &
\zeta_4(1) & 1
\end{array}\right)
\nn \\ \nn \\ \nn \\
\ldots
 \\ \nn
\ee
Since explicitly
\be
{\rm Mon}_{_{[1]}}=p_1\nn \\
{\rm Mon}_{_{[2]}}=p_2,\ \ \ \  {\rm Mon}_{_{[1,1]}}=\frac{p_1^2-p_2}{2} = {\rm Schur}_{_{[1,1]}} \nn\\
{\rm Mon}_{_{[3]}}=p_3,\ \ \ \ {\rm Mon}_{_{[2,1]}}=p_2p_1-p_3,\ \ \ \
{\rm Mon}_{_{[1,1,1]}}=\frac{p_3}{3}-\frac{p_2p_1}{2} + \frac{p_1^3}{6} = {\rm Schur}_{_{[1,1,1]}}\nn \\
{\rm Mon}_{_{[4]}}=p_4, \ \ \ \  {\rm Mon}_{_{[3,1]}}=p_3p_1-p_4, \ \ \ \
{\rm Mon}_{_{[2,2]}}=\frac{p_2^2-p_4}{2}, \ \ \ \
{\rm Mon}_{_{[2,1,1]}}=\frac{p_4}{2}-\frac{p_3p_1}{3}-\frac{p_2^2}{4} + \frac{p_1^4}{12},\nn \\
{\rm Mon}_{_{[1,1,1,1]}}=-\frac{p_4}{4} + \frac{p_3p_1}{3}
+ \frac{p_2^2}{8} - \frac{p_2p_1^2}{4} +\frac{p_1^4}{24} = {\rm Schur}_{_{[1,1,1,1]}} \nn \\
\ldots
\ee
we get:
{\footnotesize
\be
\Gns_{[1]} = p_1, \nn \\ \nn \\
\Gns_{[2]} = \left(1-\frac{\zeta_2(1)}{2}\right)p_2 + \zeta_2(1)\frac{p_1^2}{2}
= p_2 +\zeta_2(1){\rm S}_{[1,1]} = {\rm S}_{[2]} - \Big(1-\zeta_2(1)\Big){\rm S}_{[1,1]}
\nn \\
\Gns_{[1,1]} = \frac{-p_2+p_1^2}{2} = {\rm S}_{[1,1]}
\nn\\ \nn \\
\!\!\!\!\!\!\!\!\!\!\!\!
\Gns_{[3]} = \left(1-\zeta_3(1) + \frac{\zeta_3(1)\zeta_2(1)}{3}\right)p_3
+ \zeta_3(1)\left(1-\frac{\zeta_2(1)}{2}\right)p_2p_1 + \frac{\zeta_3(1)\zeta_2(1)}{6}p_1^3
= {\rm S}_{[3]} - \Big(1-\zeta_3(1)\Big){\rm S}_{[2,1]}
+ \Big(1-2\zeta_3(1)+\zeta_3(1)\zeta_2(1)\Big){\rm S}_{[1,1,1]}
\nn\\
\Gns_{[2,1]} = p_2p_1-p_3 + \Big(\zeta_2(1)+\zeta_2(t)\Big)
\left(\frac{p_3}{3}-\frac{p_2p_1}{2} + \frac{p_1^3}{6}\right)=
{\rm S}_{[2,1]} - \Big(2-\zeta_2(1)-\zeta_2(t)\Big){\rm S}_{[1,1,1]}
\nn \\
\Gns_{[1,1,1]} = \frac{p_3}{3} - \frac{p_2p_1}{2} + \frac{p_1^4}{6}={\rm S}_{[1,1,1]}
\nn
\ee
}
{\footnotesize
\be
\!\!\!\!\!\!\!\!\!\!\!\!\!\!\!
\Gns_{[4]} = p_4 + \zeta_4(1)\Big(p_3p_1-p_4\Big) +  \zeta_3(1)\zeta_3(q)\frac{p_2^2-p_4}{2} +
\zeta_3(1)\zeta_4(1)\left(\frac{p_4}{2}-\frac{p_3p_1}{3}-\frac{p_2^2}{4} + \frac{p_1^4}{12}\right)
+ \zeta_2(1)\zeta_3(1)\zeta_4(1)\left(-\frac{p_4}{4} + \frac{p_3p_1}{3}
+ \frac{p_2^2}{8} - \frac{p_2p_1^2}{4} +\frac{p_1^4}{24}\right)
\nn \\
\Gns_{[3,1]} = \Big(p_3p_1-p_4\Big) +\zeta_2(1) \frac{p_2^2-p_4}{2}
+ \zeta_2(1)\Big(\zeta_2(q)+\zeta_2(qt)\Big)
\left(\frac{p_4}{2}-\frac{p_3p_1}{3}-\frac{p_2^2}{4} + \frac{p_1^4}{12}\right)
+ \nn \\
+  \zeta_2(1)\Big(\zeta_2(1) \zeta_2(q)+\zeta_2(1) \zeta_2(qt)+\zeta_2(t)\zeta_2(qt)\Big)
\left(-\frac{p_4}{4} + \frac{p_3p_1}{3}
+ \frac{p_2^2}{8} - \frac{p_2p_1^2}{4} +\frac{p_1^4}{24}\right)
\nn \\
\Gns_{[2,2]} = \frac{p_2^2-p_4}{2}+\zeta_2(1)^2\zeta_2(qt)
\left(\frac{p_4}{2}-\frac{p_3p_1}{3}-\frac{p_2^2}{4} + \frac{p_1^4}{12}\right)
+ \zeta_2(1)\Big(\zeta_2(1)+\zeta_2(t)\Big)
\left(-\frac{p_4}{4} + \frac{p_3p_1}{3}
+ \frac{p_2^2}{8} - \frac{p_2p_1^2}{4} +\frac{p_1^4}{24}\right)
\nn\\
\Gns_{[2,1,1]} = \frac{p_4}{2}-\frac{p_3p_1}{3}-\frac{p_2^2}{4} + \frac{p_1^4}{12}
+\Big(\zeta_2(1)+\zeta_2(t)+\zeta_2(t^2)\Big)
\left(-\frac{p_4}{4} + \frac{p_3p_1}{3}
+ \frac{p_2^2}{8} - \frac{p_2p_1^2}{4} +\frac{p_1^4}{24}\right)
\nn \\
\Gns_{[1,1,1,1]} = -\frac{p_4}{4} + \frac{p_3p_1}{3}
+ \frac{p_2^2}{8} - \frac{p_2p_1^2}{4} +\frac{p_1^4}{24} \ \ \ \
\nn
\ee
}
\be
\ldots
\ee

\subsection*{Examples of conjugate polynomials $\Gns^\perp$ and their $\zeta$-grading}

The simplest of polynomials conjugate to $\Gns_R\{p\}$ w.r.t. the Schur metric
\be\label{Schumet}
\left<p_\Delta|p_{\Delta'}\right>=z_\Delta\cdot\delta_{\Delta,\Delta'},
\ \ \ \ \ \ \  \ \ \ \ \
\left<\Gns_R\{p\} \Big| \Gns^\perp_{R'}\{p\}\right> =  \delta_{R,R'}
\ee
are:
\be
\Gns_{[1]}^\perp = p_1 = {\rm S}_{[1]} \nn \\ \nn \\
\Gns_{[2]}^\perp = {\rm Schur}_{[2]} = {\rm S}_{[2]}\nn \\
\Gns_{[1,1]}^\perp = {\rm Schur}_{[1]}^2 - \zeta_2(1)\cdot{\rm Schur}_{[2]}
= {\rm S}_{[1]}^2 - \zeta_2(1)\cdot {\rm S}_{[2]}
\nn
\ee
\be
\Gns_{[3]}^\perp={1\over 6}p_1^3+{1\over 2}p_1p_2+{1\over 3}p_3={\rm S}_{[3]}\nn\\
\Gns_{[2,1]}^\perp=-{1\over 6}\Big(\zeta_3(1)-3\Big)p_1^3-{1\over 2}\Big(\zeta_3(1)-1\Big)p_1p_2
-{1\over 3}\zeta_3(1)p_3=
{\rm S}_{[1]}{\rm S}_{[2]}-\zeta_3(1){\rm S}_{[3]}\nn\\
\Gns_{[1,1,1]}^\perp=-{1\over 6}\Big(3\zeta_2(1)+3\zeta_2(t)-\zeta_2(t)\zeta_3(1)-6\Big)p_1^3+
{1\over 2}\Big(\zeta_2(1)+\zeta_2(t)-\zeta_2(t)\zeta_3(1)\Big)p_1p_2
+{1\over 3}\zeta_2(t)\zeta_3(1)p_3=\nn\\
={\rm S}_{[1]}^3-\Big(\zeta_2(1)+\zeta_2(t)\Big){\rm S}_{[1]}{\rm S}_{[2]}
+\zeta_2(t)\zeta_3(1){\rm S}_{[3]}
\nn
\ee
{\footnotesize
\be
\Gns_{[4]}^\perp={\rm S}_{[4]}\nn\\
\Gns_{[3,1]}^\perp={\rm S}_{[1]}{\rm S}_{[3]}-\zeta_4(1){\rm S}_{[4]}\nn\\
\Gns_{[2,2]}^\perp={\rm S}_{[2]}^2-\zeta_2(1){\rm S}_{[1]}{\rm S}_{[3]}-\Big(\zeta_2(1)-\zeta_2(q)\Big)\zeta_4(t)
{\rm S}_{[4]}\nn\\
\Gns_{[2,1,1]}^\perp={\rm S}_{[1]}^2{\rm S}_{[2]}-\zeta_2(1){\rm S}_{[2]}^2-\Big(\zeta_2(q)+\zeta_2(qt)
-\zeta_2(1)\Big)\zeta_2(1){\rm S}_{[1]}{\rm S}_{[3]}+
\Big(\zeta_2(q)+\zeta_2(qt)
-\zeta_2(1)\Big)\zeta_2(1)\zeta_4(t){\rm S}_{[4]}
\nn\\
\Gns_{[1,1,1,1]}^\perp={\rm S}_{[1]}^4-\Big(\zeta_2(1)+\zeta_2(t)+\zeta_2(t^2)\Big){\rm S}_{[1]}^2{\rm S}_{[2]}
+\zeta_2(1)\zeta_2(t^2){\rm S}_{[2]}^2+\nn\\
+\Big(\zeta_2(t)\zeta_2(q)+\zeta_2(t^2)\zeta_2(q)
+\zeta_2(t^2)\zeta_2(qt)-\zeta_2(t^2)\zeta_2(1)\Big)\zeta_2(1){\rm S}_{[1]}{\rm S}_{[3]}
-\Big(\zeta_2(q)+\zeta_2(qt)-\zeta_2(1)\Big)\zeta_2(1)\zeta_2(t^2)\zeta_4(t){\rm S}_{[4]}
\nn
\ee
}
\be
\ldots
\ee
Note that they are also polynomial in $\zeta$ (do not contain it in denominators)
and, more interesting, are clearly graded with the grade being related to the ordering of
Young diagrams (the grade of $\zeta_k$ is $k-1$).
From this point of view, of a special interest is level $|R|=6$,
where the ordering becomes ambiguous,
but in fact the grading rule is preserved,
because of vanishing of the coefficient in front of ${\rm Schur}_{[3]}\cdot{\rm Schur}_{[1]}^3$
in $\Gns_{[2,2,2]}$ and vice versa.
We conjecture that the same happens at higher levels.

The grading tables in the obvious notation:
{\footnotesize
\be
\!\!\!\!\!\!\!
{\rm grad}_{\zeta_2}
\left(
\Gns_{[1,1]}^\perp
 \right) \ = \
1\cdot {\rm S}_{[2]}
\nn\\
{\rm grad}_{\zeta_2}
\left(\begin{array}{c}
\Gns_{[2,1]}^\perp \\    \Gns_{[1,1,1]}^\perp
\end{array}\right) \ = \
\left(\begin{array}{cccc}
2 \\
1 & 3
\end{array}\right)
\left(\begin{array}{c}
{\rm S}_{[3]} \\
{\rm S}_{[2]}{\rm S}_{[1]}
\end{array}\right)
\nn\\
{\rm grad}_{\zeta_2}
\left(\begin{array}{c}
\Gns_{[3,1]}^\perp \\  \Gns_{[2,2]}^\perp \\ \Gns_{[2,1,1]}^\perp \\ \Gns_{[1,1,1,1]}^\perp
\end{array}\right) \ = \
\left(\begin{array}{cccc}
3 \\
4 & 1 \\
5 & 2 & 1 \\
6 & 3 & 2 & 1
\end{array}\right)
\left(\begin{array}{c}
{\rm S}_{[4]} \\ {\rm S}_{[3]}{\rm S}_{[1]} \\ {\rm S}_{[2]}^2 \\
{\rm S}_{[2]}{\rm S}_{[1]}^2
\end{array}\right)
\nn
\ee
\be
{\rm grad}_{\zeta_2}
\left(\begin{array}{c}
\Gns_{[4,1]}^\perp \\  \Gns_{[3,2]}^\perp \\ \Gns_{[3,1,1]}^\perp \\
\Gns_{[2,2,1]}^\perp \\ \Gns_{[2,1,1,1]}^\perp  \\ \Gns_{[1,1,1,1,1]}^\perp
\end{array}\right) \ = \
{\left(\begin{array}{cccccc}
4 \\
6 & 2 \\
7 & 3 & 1 \\
8 & 4 & 2 & 1 \\
9 & 5 & 3 & 2 & 1 \\
10 & 6 & 4 & 3 & 2 & 1
\end{array}\right)}
\left(\begin{array}{c}
{\rm S}_{[5]} \\ {\rm S}_{[4]}{\rm S}_{[1]} \\ {\rm S}_{[3]}{\rm S}_{[2]} \\
{\rm S}_{[3]}{\rm S}_{[1]}^2 \\ {\rm S}_{[2]}^2{\rm S}_{[1]} \\ {\rm S}_{[2]}{\rm S}_{[1]}^3
\end{array}\right), \ \ \ \ \ \ \
\ldots \ \ \ \ \ \
\nn
\ee
\be
{\rm grad}_{\zeta_2}
\left(\begin{array}{c}
\Gns_{[5,1]}^\perp \\  \Gns_{[4,2]}^\perp \\ \Gns_{[4,1,1]}^\perp \\
\Gns_{[3,3]}^\perp \\ \Gns_{[3,2,1]} \\ \Gns_{[3,1,1,1]} \\ \Gns_{[2,2,2]} \\  \Gns_{[2,2,1,1]}^\perp  \\ \Gns_{[2,1,1,1,1]}^\perp
\end{array}\right) \ = \
{\left(\begin{array}{ccccccccccc}
5 \\
8 & 3 \\
9 & 4 & 1 \\
9 & 4 & 1 \\
11 & 6 & 3 & 2 & 2 \\
12 & 7 & 4 & 3 & 3 & 1 \\
12 & 7 & 4 & 3 & 3 & 1 \\
13 & 8 & 5 & 4 & 4 & 2 & 1 & 1 \\
14 & 9 & 6 & 5 & 5 & 3 & 2 & 2 & 1 \\
15 & 10 & 7 & 6 & 6 & 4 & 3 & 3 & 2 & 1 \\
\end{array}\right)}
\left(\begin{array}{c}
{\rm S}_{[6]} \\ {\rm S}_{[5]}{\rm S}_{[1]} \\ {\rm S}_{[4]}{\rm S}_{[2]} \\
{\rm S}_{[4]}{\rm S}_{[1]}^2 \\ {\rm S}_{[3]}^2\\ {\rm S}_{[3]}{\rm S}_{[2]} {\rm S}_{[1]}
\\ {\rm S}_{[3]}{\rm S}_{[1]}^3
\\ {\rm S}_{[2]}^3 \\ {\rm S}_{[2]}^2{\rm S}_{[1]}^2 \\ {\rm S}_{[2]}{\rm S}_{[1]}^4
\end{array}\right), \ \ \ \ \ \ \
\ldots \ \ \ \ \ \
\nn
\ee
}

\bigskip

The interesting particular cases are
\begin{itemize}

\item{$\eta=0 $:
\be
\Gns_R\{p\} = {\rm Mon}_R\{p\}\ \ \ \ \ \ \ \Gns_R^\perp\{p\} = \prod_i {\rm S}_{_{R_i}}\{p\}
\ee
}

\item{$\eta=1$: this is the case of Schur polynomials,
\be
\Gns_{R}\{p\} = \Gns_{R}^\perp\{p\} ={\rm Schur}_R\{p\}
\ee
}

\item{$\eta={(1-qz)(1-t/qz)\over (1-tz)(1-z)} $: this is the case of Macdonald polynomials,
\be
\Gns_{R}\{p\} = {\rm Mac}_R\{p\},\ \ \ \ \ \ \ \Gns_{R}^\perp\{p_k\}={\rm\overline{Mac}}_{R^\vee}\{(-1)^{k+1}p_k\}
\ee
}

\end{itemize}

\bigskip

\subsection*{Examples of ring structure coefficients}

The very first examples of product decompositions for the GNS polynomials are:

{\footnotesize
\be
\Gns_{[1]}^2 = \Gns_{[2]}+ \Big(2-\zeta_2(1)\Big)\cdot\Gns_{[1,1]}
\nn \\ \nn \\
\Gns_{[1]}\cdot\Gns_{[1,1]}
= \Gns_{[2,1]} + \Big(3-\zeta_2(1)-\zeta_2(t)\Big) \cdot \Gns_{[1,1,1]} \nn \\
\Gns_{[1]}\cdot \Gns_{[2]} = \Gns_{[3]} + \Big(1+\zeta_2(1) - \zeta_2(1)\zeta_2(q)\Big)\cdot \Gns_{[2,1]}
+ \underline{\Big(2\zeta_2(1)-\zeta_2(t)-\zeta_2(1)^2-\zeta_2(1)\zeta_2(t) +
\zeta_2(1)\zeta_2(q)\zeta_2(t)\Big)} \cdot \Gns_{[1,1,1]}
\nn \\ \nn \\
\ldots \nn
\ee
}

\noindent
Note that the underlined ``anomalous" coefficient is non-vanishing for generic $\zeta_2(z)$,
but it vanishes on the Macdonald locus $\xi(z) =1-z$.

However, its conjugate counterpart vanishes for arbitrary $\xi(z)$:
\be
{\Gns_{[1]}^\perp}^2 = \Gns_{[1,1]}^\perp + \zeta_2(1)\cdot\Gns_{[2]}^\perp
\nn \\ \nn \\
\Gns_{[1]}^\perp\cdot\Gns_{[1,1]}^\perp = \Gns_{[1,1,1]}^\perp+\zeta_2(t)\cdot\Gns_{[2,1]}^\perp
\nn \\
\Gns_{[1]}^\perp\cdot\Gns_{[2]}^\perp = \Gns_{[2,1]}^\perp+\zeta_3(1)\cdot\Gns_{[3]}^\perp
\ee
The non-vanishing coefficients, which vanish on the Macdonald locus (underlined),
appear now only at level 4:
\be
\Gns_{[1]}^\perp\cdot \Gns_{[1,1,1]}^\perp = \Gns_{[1,1,1,1]}^\perp + \zeta_2(t^2)\cdot\Gns_{[2,1,1]}^\perp
\nn \\
\Gns_{[1]}^\perp\cdot\Gns_{[2,1]}^\perp = \Gns_{[2,1,1]}^\perp + \zeta_2(1)\cdot\Gns_{[2,2]}^\perp +
\zeta_2(1)\zeta_2(qt)\cdot\Gns_{[3,1]}^\perp
\nn \\
\Gns_{[1]}^\perp\cdot \Gns_{[3]}^\perp = \Gns_{[3,1]}^\perp
+ \zeta_4(1)\cdot\Gns_{[3]}^\perp
\nn \\
\Gns_{[1,1]}^\perp\cdot\Gns_{[1,1]}^\perp = \Gns_{[1,1,1,1]}^\perp+
\Big(\zeta_2(t^2)+\zeta_2(t)-\zeta_2(1)\Big)\cdot\Gns_{[1,1,2]}^\perp
+\zeta_2(1)\zeta_2(t)\cdot \Gns_{[2,2]}^\perp+\nn\\
+\underline{\Big(\zeta_2(1)^3+\zeta_2(1)\zeta_3(t)
-\zeta_2(1)^2\zeta_2(qt)-\zeta_2(1)\zeta_3(1)\Big)}\cdot\Gns_{[1,3]}^\perp
\nn \\
\Gns_{[1,1]}^\perp\cdot\Gns_{[2]}^\perp = \Gns_{[2,1,1]}^\perp +
\boxed{0}\cdot \Gns_{[2,2]}^\perp + \Big(\zeta_2(tq)+\zeta_2(q)-\zeta_2(1)\Big)\cdot\Gns_{[3,1]}^\perp
\nn \\
\Gns_{[2]}^\perp\cdot\Gns_{[2]}^\perp = \Gns_{[2,2]}^\perp + \zeta_2(1)\cdot\Gns_{[3,1]}^\perp
+ \zeta_2(q)\zeta_4(1)\cdot\Gns_{[4]}^\perp
\ee
In fact,  underlined is the only redundant coefficient at level 4.
The reason is that there are no redundant coefficients in front of $\Gns_{[r]}^\perp$ at any $r$.
The coefficient in the box is vanishing: this is in agreement with the product of representations,
$[2]\otimes[1,1] = [2,1,1]+[3,1]$, which does not contain $[2,2]$.

Note that these coefficients are obviously much simpler and better structured in the conjugate case.
This is one of the striking features of conjugate polynomials $\Gns^\perp$.

\end{document}